# Adv-Camou: An Emissivity-based Physical Dynamic Adversarial Patch for Infrared Attack

Zhuo Chen, Xu Zheng, Runyun He, Tuoyu Liu, Kaixuan Ding, Yufei Ge, Yijing Song, Tingting Shi, Haining Ji, Pinghua Tang, Haifeng Cheng, Xikun Hu, Dongqing Liu

***Abstract*—With the widespread application of infrared object detection technologies in military domains, intelligent infrared adversarial camouflage capabilities have become critical. Existing physical attack methods predominantly employ static thermal suppression strategies, suffering from limited environmental adaptability, insufficient attack robustness, and inadequate visual concealment. This study proposes Adv-Camou, an emissivity-based physical dynamic adversarial patch that achieves paradigm-shifting advancements from static to dynamic attacks through algorithm-material co-innovation. First, a multi-level loss function based on feature confusion mechanisms is designed to deceive detectors by attenuating target saliency and enforcing background feature alignment. Second, adaptive physical transformations and differentiable binarization optimization are integrated to address physical robustness and realizability constraints. A 6×4 array of reversible electrochromic emissivity device is developed and tested across a range of background scenarios, enabling scene-adaptive dynamic camouflage through independent unit control. Experiments demonstrate that this method achieves 95.59% attack success rate (ASR), reducing average precision (AP) by 60.51%. Physical validations show that static adversarial patches attain 76.55% average ASR under multi-view, multi-range, and multi-spectral conditions, while dynamic patches enhance ASR up to 91.55% in complex environments (wilderness/urban) through adaptive switching. The seamless digital-to-physical transfer via tunable emissivity devices transcends the limitations of traditional static camouflage by establishing a dynamic infrared regulation capability. This advancement enhances the battlefield survivability of military assets and demonstrates transformative potential for adaptive camouflage in practical combat scenarios.**

This work was supported in part by the Innovation Research Foundation of National University of Defense Technology (Innovation Research Foundation of NUDT), and in part by Hunan Provicial Innovation Foundation for Postgraduate. *(Corresponding author: Dongqing Liu).* Zhuo Chen and Xu Zheng are co-first authors.

Zhuo Chen, Xu Zheng, Runyun He, Tuoyu Liu, Kaixuan Ding, Yufei Ge, Yijing Song, Tingting Shi, Haifeng Cheng, and Dongqing Liu are with College of Aerospace Science and Engineering, National University of Defense Technology, Changsha, Hunan 410073, China (e-mail: 1312253843@qq.com; zhangliqiang@nudt.edu.cn; zhengxu20@nudt.edu.cn; h953725345@163.com; 1329845333@qq.com; dingkaixuan24@nudt.edu.cn). Haining Ji and Pinghua Tang are with School of Physics and Optoelectronics, Xiangtan University, Xiangtan, Hunan 411105, China. Xikun Hu is with College of Electronic Science, National University of Defense Technology, Changsha, Hunan 410073, China.

## I. Introduction

Thermal infrared object detection has emerged as a cornerstone technology in modern intelligent perception systems, spanning critical domains such as autonomous navigation, all-weather surveillance, search and rescue, and defense reconnaissance. Unlike visible-light sensors that rely on reflected ambient illumination, thermal infrared sensors capture long-wave infrared (LWIR, 8–14 μm) radiation emitted directly by objects as governed by the Stefan-Boltzmann law. Empowered by rapid advancements in deep convolutional neural networks (CNNs) and Vision Transformers (ViTs) [1][2][3], modern infrared detectors achieve remarkable precision under degraded visual conditions, including total darkness, dense fog, and heavy smoke. However, deep neural networks are inherently vulnerable to adversarial examples—inputs perturbed by imperceptible or physically structured patterns that induce high-confidence misclassifications [4][5]. Investigating physical adversarial attacks against thermal detectors is therefore imperative, both for rigorously evaluating the security boundaries of safety-critical perception pipelines and for developing next-generation intelligent stealth technologies. Over the past decade, physical adversarial attacks in the visible spectrum have been extensively investigated, evolving from localized planar patches [6][7][8] to wearable 3D-modeled garments [9][10]. Nevertheless, directly transferring visible-light adversarial paradigms to the thermal infrared domain remains fundamentally constrained by discrepancies in imaging physics. While visible-light cameras resolve high-frequency spatial textures across three chromatic channels (RGB), thermal imagers record single-channel grayscale intensity maps that reflect surface temperature and material emissivity distributions. Consequently, existing infrared physical attack methodologies predominantly rely on static thermal suppression or localized temperature manipulation [11][12][13], which encounter three fundamental bottlenecks in open-world deployment:

Lack of Environmental Adaptability: Static adversarial patches [14][15][16] feature fixed spatial topologies optimized for specific training distributions. When targets traverse heterogeneous environments—such as transitioning from structured urban streetscapes with strong mid-frequency architectural contours to cluttered wilderness terrains dominated by high-frequency foliage—the static perturbation fails to adapt to shifting background feature activations, leading to severe degradation in attack efficacy. Physical Instability and Digital-to-Physical Degradation: Approaches utilizing active heating elements (e.g., miniature bulbs [12] or resistive heating films) or phase-change temperature-controlled materials [13][17] suffer from inevitable thermal diffusion, ambient convective heat loss, and rapid power depletion. These factors blur the sharp boundaries required for high-frequency adversarial patterns and introduce uncontrollable thermal drift. Conversely, passive emissivity-based materials offer superior edge definition but can only represent discrete binary or tertiary states, creating a substantial optimization gap between continuous digital gradients and discrete physical fabrication. Limited Cross-Architecture Transferability: Many existing infrared attacks employ black-box heuristic optimizers (e.g., Particle Swarm Optimization or Differential Evolution) tailored to specific geometric primitives [18][19][20] or single-model gradient descent [21]. These methods primarily perturb shallow local responses without disrupting deep semantic representations, resulting in poor transferability across diverse detection paradigms (e.g., from anchor-based single-stage detectors to region-proposal two-stage networks and global-attention Transformers). Recent breakthroughs in nanophotonics and functional materials have enabled dynamic thermal camouflage through reversible electrochromic emissivity modulation [22][23][24]. By electrically tuning carrier densities or ion intercalation states, these devices can rapidly modulate surface thermal radiation in real time without altering actual physical temperatures. Inspired by this hardware capability, a compelling scientific question arises: Can we bridge differentiable adversarial optimization in computer vision with reconfigurable emissivity hardware in materials science to achieve scene-adaptive, dynamic physical attacks in the infrared domain? To answer this question, we propose Adv-Camou, an emissivity-based physical dynamic adversarial patch framework that establishes a paradigm shift from static thermal hiding to dynamic, environment-aware adversarial camouflage via algorithm-material co-design. In the digital domain, we formulate a multi-level feature confusion mechanism inspired by contrastive representation learning. Rather than merely minimizing output confidence scores, our loss function simultaneously suppresses intrinsic foreground saliency and aligns perturbed target features with surrounding background distributions in the latent space. To guarantee lossless digital-to-physical realizability on discrete hardware units, we introduce a differentiable binarization optimization strategy leveraging Gumbel-Softmax reparameterization [25] coupled with a Straight-Through Estimator (STE). Furthermore, we integrate complexity-aware Expectation Over Transformation (EOT) [26] and keypoint-driven perspective warping to ensure robustness against multi-view geometric distortions and long-range atmospheric attenuation. In the physical domain, we fabricate a $6 \times 4$ individually addressable array of reversible electrochromic emissivity devices, enabling real-time pattern switching across diverse operational scenarios. The main contributions of this work are summarized as follows:

Contrastive Feature Confusion Mechanism: We formulate a multi-level feature confusion loss that jointly enforces foreground feature suppression and background latent alignment, substantially enhancing attack efficacy and cross-architecture transferability across CNN- and Transformer-based infrared detectors.

Differentiable Discrete Pattern Optimization: We develop a coarse-to-fine differentiable binarization pipeline using Gumbel-Softmax relaxation and complexity-adaptive physical transformations, effectively bridging continuous gradient optimization with discrete, manufacturable emissivity unit arrays.

Algorithm-Hardware Co-Design for Dynamic Camouflage: We analyze the frequency- and scale-dependent vulnerability of infrared detectors across heterogeneous scenes and implement a $6 \times 4$ reconfigurable electrochromic emissivity array, demonstrating real-world dynamic adversarial pattern switching for the first time. Comprehensive Digital and Physical Evaluations: Extensive experiments demonstrate that Adv-Camou achieves a 95.59% attack success rate (ASR) and reduces average precision (AP) by 60.51% digitally, while physical deployments achieve up to 91.55% ASR across complex wilderness and urban environments via adaptive pattern switching. The remainder of this paper is organized as follows. Section II reviews related work in physical adversarial attacks and cross-modal infrared countermeasures. Section III details the proposed Adv-Camou methodology, including loss formulation, discrete mask optimization, and adaptive physical transformations. Section IV presents comprehensive digital evaluations, cross-model transferability analyses, and real-world static and dynamic physical experiments. Finally, Section V concludes the paper and outlines future research directions.

## II. RELATED WORKS

### *A. Physical Adversarial Attacks in the Visible Spectrum*

Physical adversarial attacks aim to generate realizable perturbations that survive real-world imaging pipelines—encompassing illumination fluctuations, viewpoint changes, and sensor noise—to deceive visual perception models. Following early demonstrations on image classifiers using printed eyeglass frames [6], localized stickers [8], and universal patches [7], research shifted extensively toward object detection, which presents greater challenges due to multi-scale regression and dense spatial anchoring. Planar patch-based methods served as the primary testbed: Yang et al. [27] leveraged the EOT framework to synthesize robust patches against YOLO detectors, while Thys et al. [28] optimized printable cardboard patches via backpropagation to evade pedestrian surveillance. To mitigate the visual conspicuousness of high-contrast perturbations, subsequent studies prioritized naturalistic stealthiness and semantic integration. For instance, Zhang and Ma [29] and Guesmi et al. [30] embedded adversarial perturbations into semantic contexts and artistic textures, respectively, whereas Zhou et al. [31] balanced perceptual realism with deception strength. Generative Adversarial Networks (GANs) and manifold projections have also been widely adopted to synthesize naturalistic patches [32], [33], alongside Principal Component Analysis (PCA) to construct eigenpatches with superior cross-model transferability [34]. Because planar patches fail under severe viewing angles and non-rigid human motion, wearable adversarial camouflage has emerged as a prominent research trajectory. Xu et al. [9] introduced Thin Plate Spline (TPS) transformations to model the non-rigid wrinkling of adversarial T-shirts, while Huang et al. [35] jointly optimized classification and regression branches for universal camouflage. To achieve omnidirectional evasion, Wu et al. [36] and Li et al. [37] developed multi-view optimization frameworks for wearable garments. Pushing the boundaries of spatial coverage, Hu et al. proposed toroidal-cropped repeatable textures (AdvTexture) [10] and 3D-modeled voronoi clothing textures (AdvCaT) [38] to ensure full-angle physical realizability. More recently, Cheng et al. [39] introduced a block-decoupling strategy (DePatch) to alleviate feature coupling interference during physical deployment. Despite these advances, visible-light methods rely exclusively on manipulating high-resolution RGB reflectance, rendering them ineffective against thermal infrared sensors that operate in complete darkness and perceive radiative emission rather than surface color.

### *B. Thermal Infrared and Cross-Modal Physical Attacks*

To counter thermal infrared detectors, physical perturbations must manipulate either surface temperature or material emissivity inside the LWIR spectrum. Early pioneering efforts by Zhu et al. constructed active thermal perturbations using arrays of miniature incandescent bulbs [12] and QR-code-like thermal insulation clothing [17] to deceive pedestrian detectors. However, active power sources and complex wiring limit wearability and long-term stability. Consequently, passive patches utilizing low-emissivity metals or thermal insulation tapes gained traction. Because the shapes and placements of physical materials are often non-differentiable, zero-order heuristic algorithms have been widely utilized. Wei et al. [11] employed Particle Swarm Optimization (PSO) to determine the placement of warm and cold stickers, while Hu et al. applied Differential Evolution (DE) and PSO to optimize thermal blocks (AdvIB) [13] and Bézier-curved strips (AdvIC) [18]. Similarly, Tiliwalidi et al. explored geometric primitives (AdvIG) [19] and multi-view grid structures (AdvGrid) [15] via heuristic search. While avoiding gradient computation, these heuristic approaches restrict the perturbation space to simplistic geometric shapes, lacking the structural complexity needed to disrupt deep feature pyramids in cluttered backgrounds. To incorporate gradient information and address multi-sensor fusion systems, white-box shape optimization and visible-infrared cross-modal attacks have recently been explored. Wei et al. proposed learnable infrared patch shapes (AIP) [21] and unified cross-modal boundary patches (UCA) [14] to simultaneously fool RGB and thermal detectors. Building upon this, Hu et al. introduced two-stage cross-modal optimization [20] and bio-inspired leaf-structured patches [16], while Zhu et al. [40] designed 3D-aware infrared car stickers. Nevertheless, existing infrared and cross-modal patches remain fundamentally static after physical fabrication. Fine-grained stripe or grid structures [15], [17] suffer from severe spatial aliasing at extended distances due to the low spatial resolution of thermal focal plane arrays, and static topologies cannot adapt to dramatic shifts in background thermal distributions. In contrast to all prior static schemes, our proposed Adv-Camou integrates differentiable binarized optimization with an electrically tunable electrochromic emissivity array, enabling scene-adaptive dynamic switching and robust multi-scale feature confusion in complex physical environments.

## III. Methodology

### *A. Loss Function Design*

In infrared object detection tasks, the primary objective of adversarial attacks lies in generating physically deployable adversarial patches that disrupt the feature extraction process of detection models. This interference significantly reduces the target's feature response in thermal imaging, thereby achieving thermal invisibility through feature suppression. Formally, considering an original infrared image $I_0 \in \mathbb{R}^{H\times W}$ containing pedestrian targets as input to a trained detection model $f(\cdot)$, the model processes the input to produce predictions $f(I_o) \to L_{pre}$, which are compared against ground truth annotations $L_C$. The prediction output $L_{pre} := \left[V_{pos}, V_{obj}, V_{cls}\right]$ comprises three critical components: positional coordinates $V_{pos} = (b_x, b_y, b_w, b_h)$ representing the detected bounding box's center coordinates ($b_x, b_y$) and dimensions ($b_w, b_h$), confidence score $V_{obj}$ for target presence, and categorical probability $V_{cls}$ for class identification.

$$L_{pre} := \left[V_{pos}, V_{obj}, V_{cls}\right] = f(I_o) \tag{1}$$

The adversarial patch is formally defined as $\delta \in [0,1]^{c\times h\times w}$, where $h$ and $w$ denote the patch dimensions constrained by $0<h<b_h$ and $0<w<b_w$, respectively, ensuring physical proportionality to target bounding boxes. A perturbation mask $M \in \{0,1\}^{1\times h\times w}$ governs patch application through element-wise multiplication, generating adversarial inputs via:

$$I_o' = \left(1 - M\right) \odot I_o + M \odot \delta \tag{2}$$

This operation induces model prediction errors:

$$y^{true} \neq \arg\max f\left(I_o'\right) \tag{3}$$

As illustrated in Fig. 1(b), the detector's feature extraction module produces $C$-channel feature maps with dimensions $v_h \times v_w$. Bounding box coordinates undergo proportional scaling to derive feature-level dimensions ($H$ (anchor) and $W$ (anchor)):

$$vb_h = \frac{b_h \cdot v_h}{H}, vb_w = \frac{b_w \cdot v_w}{W} \tag{4}$$

The foreground feature vectors $V_{1:c}^{fore}$ within the bounding box region of feature maps encapsulate target-specific characteristics, while the background feature vectors $V_{1:c}^{back}$ outside this region capture contextual information. This work establishes a concealment-oriented adversarial attack framework grounded in feature confusion mechanisms for detection evasion. The attack objective is decomposed into dual feature-space operations: 1) attenuation of target saliency representations to induce misclassification as background noise, and 2) alignment of target feature vectors with background distributions in latent space. This adversarial mechanism exhibits topological isomorphism with contrastive learning paradigms, where canonical target features serve as anchors, background features as positive samples, and adversarial targets as negative samples. Optimal attack efficacy is achieved through simultaneous minimization of anchor-positive distance and maximization of anchor-negative divergence, thereby realizing adversarial feature decoupling.

The loss function combines both background feature alignment term ($L_{fb}$) and suppression components ($L_{ff'}$):

$$L = L_{fb} + L_{ff'} \tag{5}$$

The adversarial pattern is optimized through iterative backpropagation that minimizes the total loss $L$.

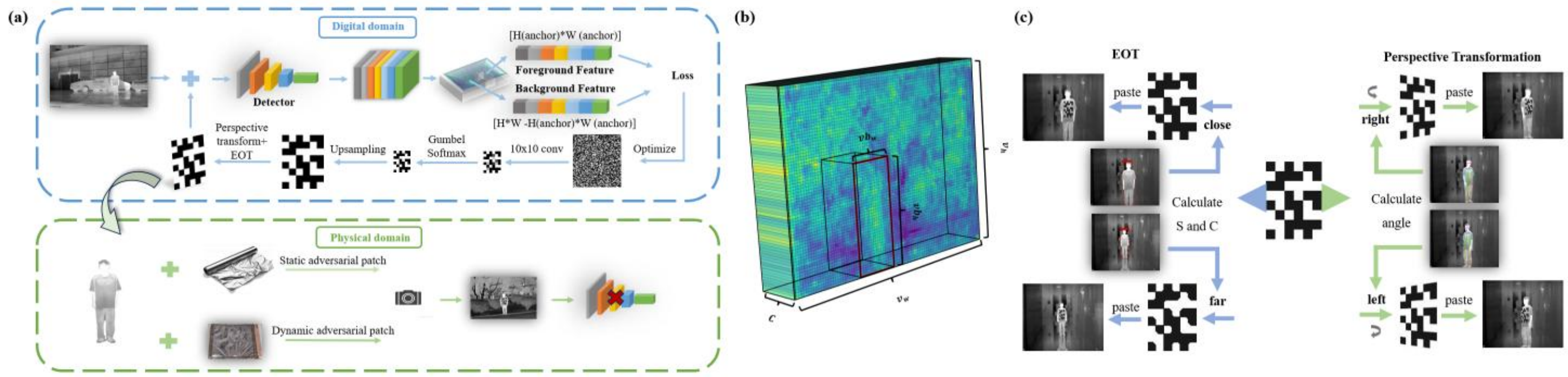

**Fig. 1.** Framework for Physical Attacks Using Adversarial Camouflage Patches. (a) Implementation workflow of adversarial camouflage patch attacks across digital and physical domains. The digital domain involves generating adversarial patterns through feature confusion loss optimization and Gumbel-Softmax-based binarization, while the physical domain focuses on deploying these patterns using reconfigurable emissivity devices. (b) Visualization of image feature vector combination. (c) Schematic diagram of physical simulation transformations for adversarial camouflage patterns.

1) **Background Feature Alignment Term**

The adversarial objective requires minimizing feature similarity between the perturbed target region and background areas to induce misclassification of target features as background components. This is formalized through the following loss function:

$$L_{fb} = -\frac{1}{N_{fb}} \sum_{i=1}^{N_{fb}} \log(1 - S_{fb}) \tag{6}$$

where the similarity comparison count $N_{fb}$ is computed by:

$$N_{fb} = vb_h \cdot vb_w \cdot (v_h \cdot v_w - vb_h \cdot vb_w) \tag{7}$$

The minimized pairwise similarity metric $S_{fb}$ is defined as:

$$S_{fb} = \operatorname{sim}\min(V_{1:c}^{fore}, V_{1:c}^{back}) \tag{8}$$

Here, $N_{fb}$ denotes the total number of feature similarity comparisons between perturbed target and background regions, while sim min(·) represents the similarity minimization operation.

2) **Target Feature Suppression Term**

The optimization aims to maximize feature similarity between perturbed and original target regions, thereby disrupting the intrinsic feature distribution to prevent reliable target characterization. The suppression loss function is formulated as:

$$L_{ff'} = -\frac{1}{N_{ff'}} \sum_{i=1}^{N_{ff'}} \log(S_{ff'}) \tag{9}$$

where the comparison count $N_{ff'}$ is derived from:

$$N_{ff'} = (vb_h \cdot vb_w)^2 \tag{10}$$

The maximized pairwise similarity metric $S_{ff'}$ is defined as

$$S_{ff'} = \operatorname{sim}\max(V_{1:c}^{fore_ori}, V_{1:c}^{fore_adv}) \tag{11}$$

Here, $S_{ff'}$ enumerates feature similarity comparisons between original $V_{1:c}^{fore_ori}$ and adversaria $V_{1:c}^{fore_adv}$ target regions, with sim max(·) denoting the similarity maximization operator.

*B. Adversarial Mask Patterns Optimization*

To address the physical deploy ability requirements of infrared adversarial masks, a differentiable binarized pattern generation method is proposed. Since infrared imaging directly maps grayscale values to thermal distributions, adversarial patches must be designed as discrete binary structures: black regions (low-emissivity materials) and white regions (normal skin temperature). Conventional gradient-optimized masks $M \in [0,1]$ produce continuous values unsuitable for physical fabrication. This limitation is resolved through Gumbel-Softmax [32] reparameterization, where each pixel's binarization probability is modeled as $\pi_0$ (probability of 0) and $\pi_1$ (probability of 1), constrained by $\pi_0+\pi_1=1$.

The Gumbel noise generation process is defined as:

$$g_i = -\log(-\log(U_i)) \qquad U_i \sim Uniform(0,1) \tag{12}$$

The sampling vectors $y_i$ ($i$=0,1) are computed to approximate one-hot encoding through temperature-controlled relaxation. Let $\tau$ denote the temperature hyperparameter, where lower $\tau$ values (e.g., $\tau$=0.1 in experiments) yield sharper approximations to discrete one-hot vectors:

$$y_i = \mathrm{Softmax}[(g_i + \log \pi_i)/\tau)] \tag{13}$$

For grayscale pixels normalized to [0,1], the binarized approximation at each pixel is derived as $p^* \approx y_0 \cdot 0 + y_1 \cdot 1 = y_1$. This differentiable approximation enables gradient-based optimization of $\pi_i$ while maintaining discrete constraints. A Straight-Through Estimator (STE) is integrated to bypass non-differentiable thresholding during backpropagation, where forward passes apply hard binarization ($p^*\geq 0.5\rightarrow 1$, otherwise 0).

To mitigate local optima risks, structural coherence is enhanced through a coarse-to-fine strategy: Apply 10×10 average pooling to an initial 80×60 noise map, generating an 8×6 coarse-grained mask; Restore full resolution via bilinear up sampling; Refine edges through Gaussian blur regularization. Experimental results demonstrate that the Gumbel-Softmax framework effectively bridges continuous optimization with discrete binarization requirements for infrared-adversarial pattern generation.

### *C. Physical Simulation Transformation*

To ensure adversarial patterns maintain robust performance across real-world deployment scenarios, the optimization framework integrates EOT [26] and perspective transformations during digital simulation. This methodology synthesizes physical-world environmental factors (e.g., lighting variations, thermal drift) and multi-view observation angles to bridge the simulation-to-reality gap. The effectiveness of this approach is demonstrated in Fig 1(c).

#### 1) Adaptive EOT Enhancement

Conventional fixed-parameter EOT methods exhibit limited robustness for adversarial patches in complex infrared scenarios involving background thermal interference, multipath reflections, and long-range imaging degradation. To address this limitation, a dynamic EOT transformation strategy is proposed, which adaptively adjusts data augmentation intensity based on quantified environmental complexity C and target bounding box S dimensions.

Environmental complexity is predicted by a dual-branch Interpretable Complexity Network (ICNet), incorporating a Spatial Layout Attention Module (SLAM) for multiscale feature modeling. Trained on the IC9600 dataset comprising 9,600 annotated infrared images, ICNet [41] achieves effective scene discrimination (outdoor environments exhibiting significantly higher complexity scores than indoor settings). The relative target size $S$ is calculated as $S=b_h/h$ where $b_h$ denotes detected bounding box height and $H$ represents image height.

$$w = \alpha \cdot \frac{C_{max}}{C} + \beta \cdot (1 - \frac{S_{max}}{S}) \tag{14}$$

where $C_{max}=S_{max}=1$ denote normalized upper bounds, with $\alpha$ and $\beta$ as tunable coefficients. This formulation drives parameter-adaptive EOT transformations: high-complexity scenarios ($C\rightarrow 1$) receive intensified noise injection and thermal drift simulation, while distant targets ($S\rightarrow 0$) undergo Gaussian blurring to mimic atmospheric attenuation effects.

#### 2) Adaptive Perspective Transformation

To achieve dynamic alignment between adversarial patches and human targets, a pose-aware perspective transformation method is developed. Body keypoints — left/right shoulder $(x_{ls},y_{ls})$, $(x_{rs},y_{rs})$ and left/right hip $(x_{lh},y_{lh})$, $(x_{rh},y_{rh})$ — are extracted using the OpenPose [42] framework to determine patch placement regions. The initial patch corners $A(0,0)$, $B(w,0)$, $C(w,h)$, and $D(0,h)$ (where $w$ and $h$ denote patch width and height) undergo perspective warping based on anatomical keypoints.

Given source patch corners and target keypoint coordinates, a homography matrix $H \in \mathbb{R}^{3\times 3}$ is computed to align the patch with the target surface:

$$\begin{bmatrix} wx' \\ wy' \\ w \end{bmatrix} = H \begin{bmatrix} x \\ y \\ 1 \end{bmatrix} \tag{15}$$

Where $H = \begin{bmatrix} h_{11} & h_{12} & h_{13} \\ h_{21} & h_{22} & h_{23} \\ h_{31} & h_{32} & h_{33} \end{bmatrix}$ defines the projective mapping between the patch plane (source coordinates $(x,y)$) and body surface (target coordinates $(x',y')$), with $w$ as the homogeneous scaling factor. The optimal $H$ is derived via least squares estimation by solving the overdetermined system constructed from corresponding corner-keypoint pairs. This transformation ensures precise geometric alignment of adversarial patterns with dynamically varying human poses.

## III. Experiments

### *A. Adversarial Camouflage Patch Generation*

Based on the FLIR ADAS v1.3 dataset, an 8×6 binary patch (0: low-emissivity regions, 1: normal-temperature regions) was applied to the upper body of pedestrians. The feature confusion loss was minimized using a momentum SGD optimizer (batch size: 8, initial learning rate: 0.05). As shown in Fig. 2(a), the generated adversarial patch achieved an ASR of 81.18% against YOLOv5 on the test set, with the AP dropping sharply from 98.28% to 37.77% (a reduction of 60.51%), effectively inducing missed detections (For more details, see *Supplementary Material* Note 1).

The adversarial patch optimized by posture adaptation showed stronger attack effectiveness on the self-made dataset, with an overall ASR of 95.59%. Specifically, the attack performance can still be maintained in extreme side view angles and distant scenes, demonstrating the strong robustness of this method to changes in perspective and scaling. (For more details, see *Supplementary Material* Note 2).

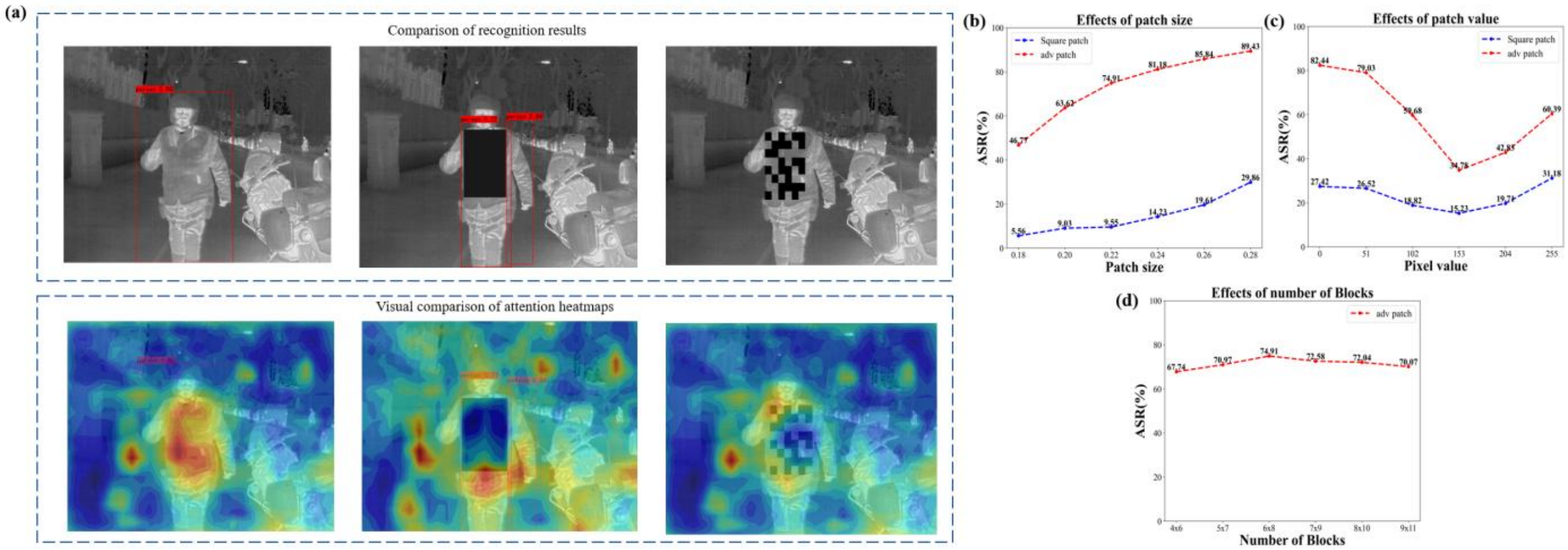

**Fig. 2.** Examples and Comparative Test Results of Adversarial Camouflage Patch Attacks. (a) Comparison of detection effects and feature attention heatmaps among the original image, an image with a rectangular patch, and an image with an adversarial camouflage patch. (b) ASR test results for patches of varying sizes. (c) ASR test results for patches with different pixel configurations. (d) ASR test results for patches with varying resolutions.

To optimize attack performance, this study systematically evaluated the impact of patch size, grayscale values, and resolution. A black square patch was used as a baseline to validate the effectiveness of the proposed Adv-Camou patches. Experimental results demonstrated that increasing the relative patch size from 0.18 to 0.28 linearly improved the ASR by 42.66% (from 46.77% to 89.43%), as shown in Fig. 2(b). This enhancement stems from larger patches covering broader thermal signature regions, intensifying feature confusion. Notably, under identical size constraints, Adv-Camou achieved 41% higher ASR than the baseline black patch, underscoring the necessity of structural optimization.

Grayscale values significantly influenced thermal contrast: extreme values (0 or 1) maximized ASR (82.44% for grayscale 0 and 60.39% for grayscale 1, Fig. 2(c)), while intermediate values (e.g., 0.6) reduced ASR to 34.78% due to similarity with background grayscale ($\approx 0.7$). This highlights high contrast as a critical factor for effective attacks.

Resolution analysis revealed that an 8×6 grid balanced texture complexity and optimization stability, achieving 74.91% ASR (Fig. 2(d)). Lower resolutions (6×4) degraded ASR by 7.17% due to oversimplified patterns, while higher resolutions (11×9) suffered from local optima issues, reducing ASR by 18%. These findings indicate that patch size and contrast dominate attack strength, while resolution governs pattern optimizability. Through parameter co-optimization (recommended: size $\geqslant 0.22$, grayscale 0, $8 \times 6$ resolution), robust attacks with ASR >80% can be achieved in complex infrared scenarios.

### *B. Static Adversarial Camouflage Patch Attack*

To validate the robustness of Adv-Camou, we compared it with baseline methods—Random Patch, QR Code, AIP, and UCA—under identical perturbation regions. As summarized in Table 1, Adv-Camou achieved an ASR of 81.18% on the test set, significantly outperforming AIP (23.66%) and UCA (21.32%). The AP of YOLOv5 plummeted from 98.28% to 37.77% (a 60.51% reduction), while comparative methods retained AP values above 89% (For more details, see *Supplementary Material* Note 3). The superior performance of Adv-Camou arises from its multi-scale feature alignment mechanism, which generates structurally complex patterns to disrupt the multi-scale feature fusion process of detection models.

In contrast to QR Code-based attacks, Adv-Camou's sparse binarized structure ensures precise coverage of critical feature attention regions (e.g., shoulders and torso), effectively dismantling the model's holistic perception of human anatomy (For more details, see *Supplementary Material* Note 3). QR codes, however, suffer from high-frequency texture distortion at long ranges, while AIP/UCA only locally suppress features without blocking feature propagation pathways. Further tests on a custom dataset (multi-view, multi-distance) demonstrated Adv-Camou's exceptional adaptability: it achieved 95.59% ASR (Fig. 3), maintaining robust performance under extreme side views (±45°) and long distances (>10 m). In overhead angle experiments (45°/60°), geometric adaptation enabled the patch to counteract target deformations, sustaining precise coverage of key thermal regions under dynamic conditions. These results confirm that Adv-Camou's structural optimization and feature confusion mechanisms significantly enhance attack robustness in complex surveillance scenarios.

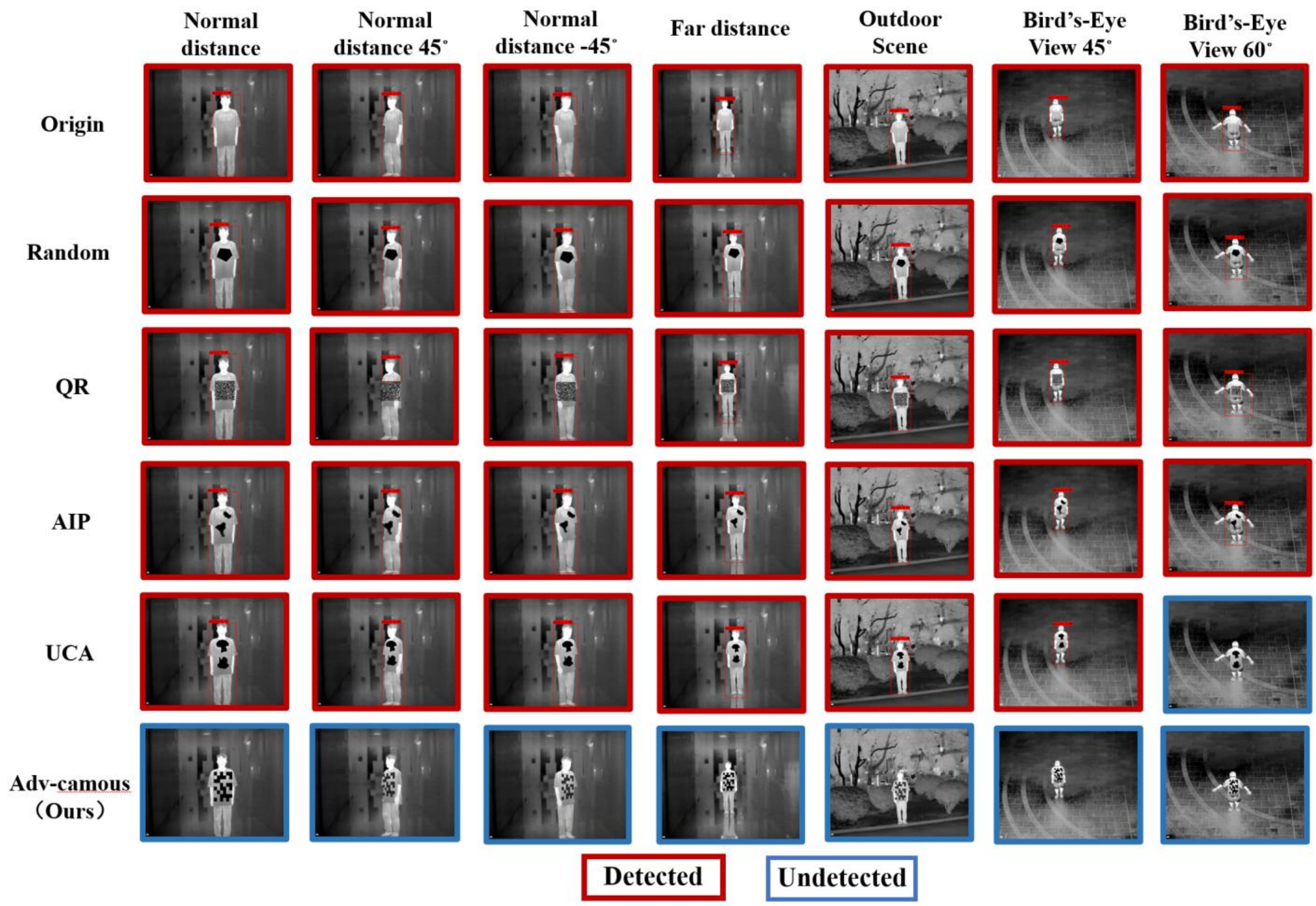

**Fig. 3.** Multidimensional Attack Efficacy Comparison Between Adv-Camou and Other Methods. "Origin" : original unprocessed images; "Random" : images with randomly placed patches; "QR" : images with patches generated by the QR code-based adversarial attack method[17]; "AIP" : the Advanced Infrared Patch attack method[21]; "UCA": the Unified Cross-modal Attack method[14].

TABLE I
COMPARISON TEST RESULTS OF DIFFERENT ATTACK METHODS

| Methods | YOLO v5 | |
|---|---|---|
| | AP↓ | ASR↑ |
| Origin | 98.68% | - |
| Random | 91.30% | 5.43% |
| QR[17] | 89.88% | 15.78% |
| AIP[21] | 89.80% | 23.66% |
| UCA[14] | 90.44% | 21.32% |
| Adv-Camou | 37.77% | 81.18% |

To evaluate the generalization capability of Adv-Camou, we tested its attack efficacy on six mainstream detectors, including single-stage, two-stage, and Transformer-based architectures. Experimental results (Table II) demonstrate that Adv-Camou achieved an ASR of 65.81% and reduced the mean AP by 59.22% (For more details, see *Supplementary Material* Note 4). Single-stage models (YOLOv5, SSD, RetinaNet), which rely on dense anchor mechanisms, exhibited high vulnerability with ASR exceeding 80% and AP dropping below 25% (e.g., RetinaNet: AP=16.41%, ASR=85.83%). For two-stage models, Faster R-CNN mitigated interference through its Region Proposal Network (RPN), limiting ASR to 59.57% (AP=14.36%). The Transformer-based DETR, leveraging global self-attention to resist feature perturbations, showed greater resilience with ASR of 49.54% (AP=44.93%). YOLOv8 localized attack impacts via multi-scale feature pyramid fusion, achieving ASR=23.95% (AP=77.15%), underscoring the critical role of model architecture in robustness.

Notably, single-stage models exhibit greater susceptibility to localized perturbations owing to the direct coupling between features and anchors. In contrast, two-stage and transformer-based architectures afford partial defense against attacks via cascaded optimization or global attention mechanisms, yet they still retain vulnerabilities in the localization task. These findings confirm the broad applicability of Adv-Camou across diverse detection frameworks, highlighting its adaptability to both anchor-driven and attention-based paradigms.

TABLE II
COMPARISON TEST RESULTS OF DIFFERENT ATTACK METHODS

| Models | Ori | | Camou Block | |
|---|---|---|---|---|
| | Ap↓ | ASR↑ | Ap↓ | ASR↑ |
| YOLO v5 [43] | 98.68% | - | 37.77% | 81.18% |
| Faster RCNN [1] | 93.75% | - | 14.36% | 59.57% |
| YOLO v3+Eff [44] | 95.38% | - | 37.19% | 78.59% |
| YOLO v8 [45] | 93.83% | - | 77.15% | 23.95% |
| RetinaNet [2] | 92.08% | - | 16.41% | 85.83% |
| SSD [46] | 92.30% | - | 21.96% | 82.02% |
| DETR [3] | 98.29% | - | 44.93% | 49.54% |

A multidimensional experimental framework was designed to comprehensively evaluate adversarial camouflage patches' effectiveness under real-world variables including distance, viewing angle, and complex backgrounds. With the detection confidence threshold set at 0.3, baseline tests confirmed 98.7% recall for unperturbed targets within 2-8 meters and ±45° horizontal angles, validating the detector's high reliability in conventional scenarios.

The adversarial evaluation employed a multimodal dataset: distance-axis sampling at 2-meter intervals (2 m/4 m/6 m/8 m) with tri-view acquisition (-45°/0°/+45°) per distance, yielding 24 video sequences (100 frames each). Experimental results (Fig. 3) demonstrate 94.45% average ASR across scenarios, peaking at 98.2% for close-range (2 m) attacks and maintaining 90.1% efficacy at 8 m. Extreme lateral angles (±45°) achieved 91.6% ASR, confirming Adv-Camou's effectiveness in compensating for geometric distortions. Further validation in complex outdoor environments revealed context-dependent performance: 82.42% ASR in the field background (vegetation clutter) versus 52.78% in the urban background (structured). Though environmental thermal noise degrades attack consistency, the method sustains practical effectiveness across heterogeneous condition (Fig. 4).

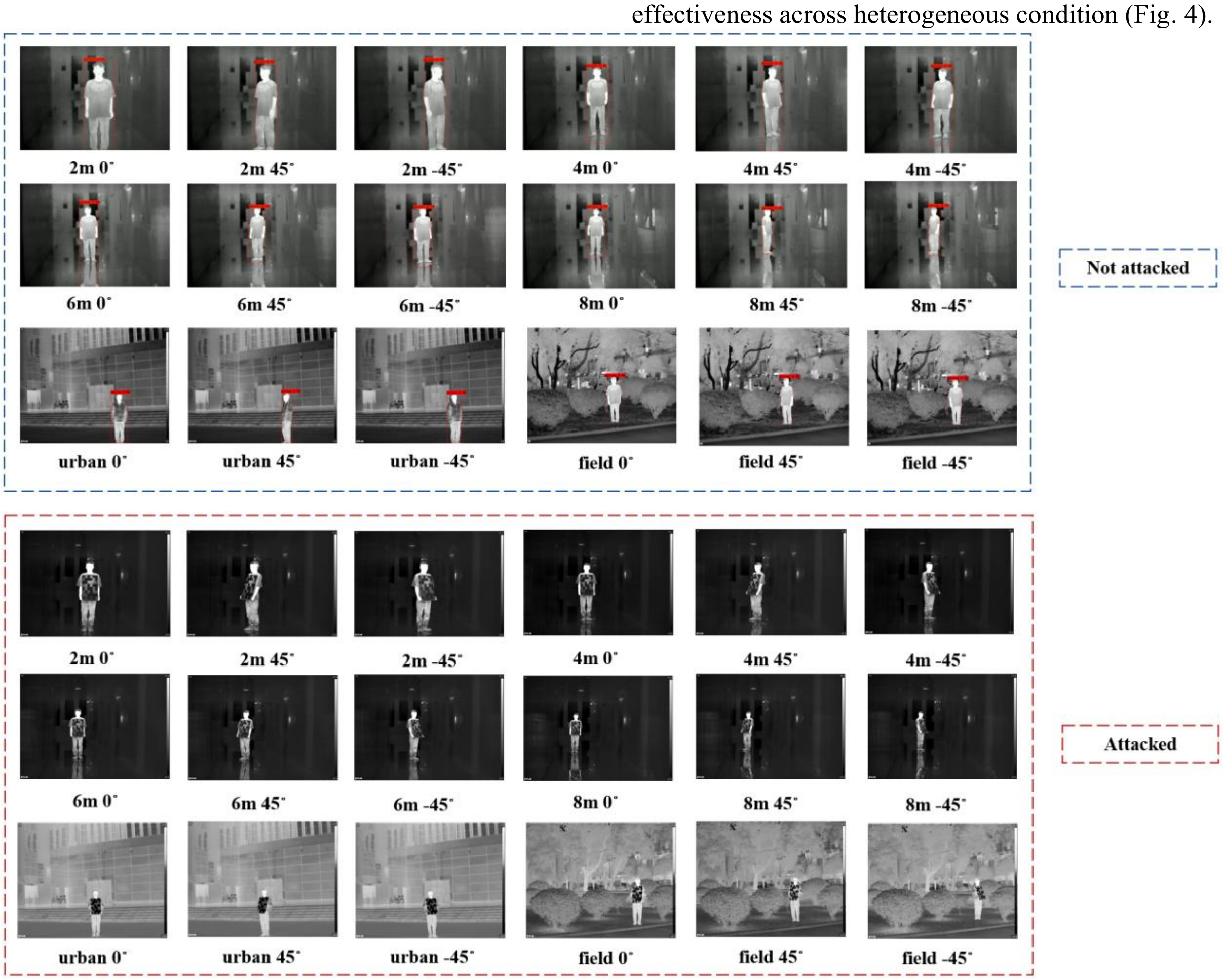

**Fig. 4.** Effectiveness of physical adversarial camouflage attacks. The blue box represents the actual person target sampling detection effect, and the red box represents the detection effect after pasting physical adversarial camouflage patches on the actual person target.

These findings highlight Adv-Camou's physical-world viability, particularly its robust performance across extended operational ranges (2-8 m) and multiview angles (0°-45°). The results provide empirical support for deploying adversarial patches in real-world surveillance systems, demonstrating resilience against geometric variations and environmental noise through coordinated pattern-structure optimization.

### *C. Dynamic Adversarial Camouflage Patch Attack*

Investigations into infrared adversarial attacks under complex environments reveal significant scene-dependent variations in the feature attention patterns of detection models. In the urban background, structured architectural reflections disperse attention to background edges, while the field background exhibit localized attention drift triggered by high-frequency textures (e.g., foliage). Cross-scenario attack evaluations were conducted using the 8×6 adversarial camouflage matrix, leading to context-aware patch strategies: Urban Adv-Camou suppresses low-frequency contour features for urban environments, whereas Field Adv-Camou disrupts mid/high-frequency textures for wilderness conditions.

Experimental results on YOLOv5 demonstrate dynamic patch-switching efficacy: urban-optimized patches reduce confidence scores from 0.42 to 0.13 in cityscapes, while wilderness-adapted patches decrease scores from 0.39 to 0.21 in natural settings. (Fig. 5(a)).

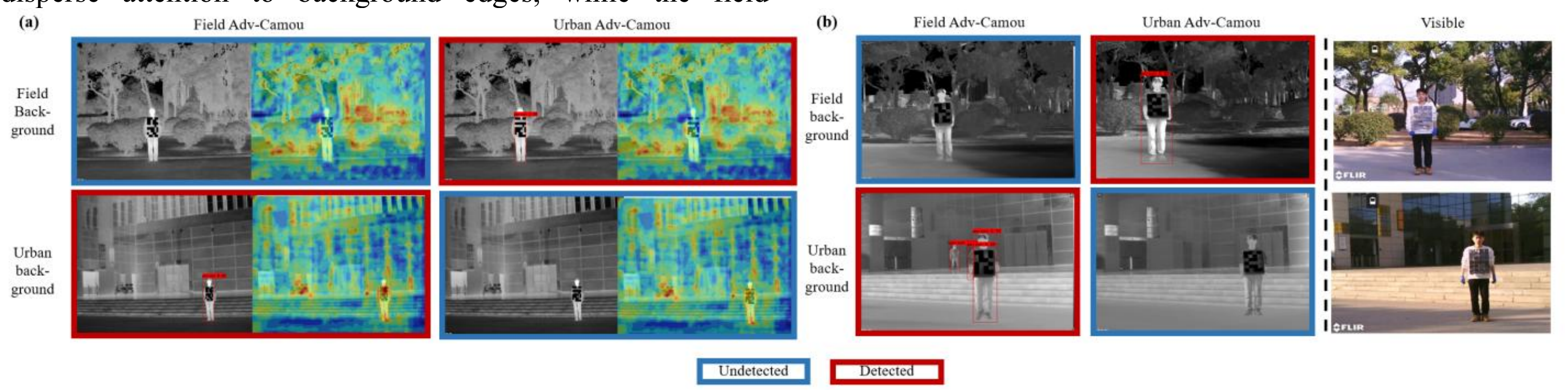

**Fig. 5.** Dynamic Adversarial Camouflage Patch Attack Performance. Blue bounding boxes: successful evasion of detection, red boxes: detected targets. (a) Detection results and corresponding feature attention heatmaps for digital attacks using different adversarial camouflage patterns across distinct scenarios. (b) Detection results and visible-light imaging effects of physical attacks with scene-specific adversarial patches.

Mechanistic analysis uncovers three interference principles. First, shallow CNN kernels predominantly capture local textures (edges, corners). High-frequency environmental textures (leaf veins, architectural joints) activate redundant low-level channels, competing with adversarial perturbations for critical feature modulation. Second, deep layers with large receptive fields extract environmental structural semantics. Urban geometric regularities (parallel lines, right angles) establish spatial priors, making targets vulnerable when patches disrupt geometric consistency. Third, environmental textures dominantly activate specific spectral bands—wilderness high-frequency vs. urban mid-frequency channels. Effective attacks require adversarial patterns to overlap targets' sensitive frequency ranges; Feature Pyramid Networks (FPNs) exhibit semantic conflicts between small-scale (branches) and large-scale (buildings) features in complex environments, weakening patch-induced multi-scale perturbations.

These findings confirm that dynamic scene adaptation overcomes universal adversarial patch limitations, achieving optimal interference amidst multi-scale feature conflicts through environment-aware frequency targeting and structural disruption.

TABLE III
PHYSICAL ATTACK RESULTS OF ADVERSARIAL CAMOUFLAGE PATCHES IN DIFFERENT SCENARIOS

| Scenarios | Inside （ASR） | Outside（ASR） | |
|---|---|---|---|
| | | Field background | Urban background |
| General Adv Camou | 94.45% | 82.42% | 52.78% |
| Field Adv Camou | 93.26% | 91.18% | 31.03% |
| Urban Adv Camou | 93.68% | 68.57% | 80.43% |

As summarized in Table III, physical experiments with static adversarial patches (aluminum foil-insulating tape composite structures) achieved an ASR of 94.45% (indoor), 82.42% (field), and 52.78% (urban), with urban performance degradation attributed to complex thermal background interference. To address this, a dynamic scene-adaptive strategy was developed: Field Adv-Camou suppresses mid-high frequency noise for wilderness textures, while Urban Adv-Camou disrupts low-frequency structural patterns in urban environments (Fig. 5(b)). Experimental results reveal significant ASR improvements when patches match operational contexts: Field Adv-Camou attains 91.18% ASR in wilderness (28.4% improvement over generic patches), and Urban Adv-Camou achieves 80.43% ASR in the urban background (27.65% gain). Conversely, context mismatch (e.g., Field patches in cities) collapses ASR to 31.03% (60.15% reduction) due to frequency-band misalignment (For more details, see *Supplementary Material* Note 5).

The experimental results verify that the dynamic patch-switching strategy can restore attack effectiveness and effectively improve the ASR in cross backgrounds, achieving maximum interference energy in complex physical environments. This mechanism can adapt to dominant features in different scenarios and overcome semantic conflicts in feature fusion.

The Adv-Camou methodology generates 6×4 adversarial camouflage patterns implemented via a tunable emissivity array (24 individually addressable elements, see *Supplementary Material* Note 8). Voltage modulation (±2.5 V) enables dynamic emissivity control ($\Delta\varepsilon$=0.32) for scene-adaptive camouflage. As summarized in Table IV, experimental results demonstrate 82.35% ASR for Field Adv-Camou in wilderness environments versus 68.63% for Urban Adv-Camou. Conversely, Urban Adv-Camou achieves superior urban performance (56.25% ASR) compared to Field Adv-Camou (41.67%). A dynamic patch-switching strategy enhances cross-scenario adaptability, yielding 13~15% average ASR improvement and validating the superiority of dynamic physical attacks.

Long-distance testing (>15 m) confirms consistent adversarial performance patterns under spatial scaling (The example sample images are shown in *Supplementary Material* Note 6). Thermal imaging analysis reveals stable thermal differentials of 2-3°C between low-emissivity (black) and high-emissivity (gray) regions, ensuring sustained stealth and operational persistence. Experimental results demonstrate that coordinated optimization of dynamic emissivity devices and adversarial algorithms achieves lossless digital-to-physical transfer across varying observation distances. This hardware-algorithm synergy enables robust deployment of adaptive camouflage systems in real-world infrared surveillance scenarios.

TABLE IV
PHYSICAL ATTACK RESULTS OF DYNAMIC ADVERSARIAL CAMOUFLAGE PATCH

| Scenarios | Field background（ASR） | Urban background（ASR） |
|---|---|---|
| Field Adv-Camou | 82.35% | 41.67% |
| Urban Adv-Camou | 68.63% | 56.25% |

## IV. CONCLUSION

This study proposes Adv-Camou, a novel infrared adversarial camouflage patch attack method capable of evading infrared target detectors in complex physical environments while generating robust adversarial patterns. The Adv-Camou framework first extracts feature maps from a target detection model and designs a loss function to generate aggressive patches. Subsequently, the Gumbel-Softmax technique is employed to binarize patch pixels, ensuring physical realizability. Finally, an adaptive physical transformation method is provided to enhance the robustness of adversarial patches in real-world conditions.

Experimental results validate the transferability of Adv-Camou, demonstrating its effectiveness against most mainstream infrared detectors. Compared to existing adversarial attack methods, Adv-Camou exhibits superior performance in the ASR and cross-model adaptability, proving its potent attack capabilities in infrared detection systems.

For physical implementation, aluminum foil and insulating tape were utilized as static emissivity materials, deployed on human targets from multiple angles. Comprehensive evaluations confirm that Adv-Camou successfully evades infrared pedestrian detection systems across varying distances, angles, and environments. Notably, in multi-scenario and multi-distance experiments, Adv-Camou demonstrates exceptional attack efficacy and adaptability.

Furthermore, using static patches generated by Adv-Camou, this study analyzes the impact of environmental features on adversarial attacks and validates the advantages of dynamic patch-switching strategies. By implementing electrochromic emissivity devices, dynamic physical attacks were achieved in diverse environments. Results show that dynamic patches significantly outperform static ones, indicating that modulating emissivity dynamically enhances ASR and adaptability, particularly in complex scenarios.

To conclude, this research establishes a multi-angle, high-success-rate adversarial attack framework against infrared pedestrian detectors in real-world settings, showcasing Adv-Camou's robustness across environments and devices. However, this study has some limitations: (1) Experiments focused on single-target scenarios, neglecting multi-target or group detection contexts; (2) Adv-Camou is designed for long-wave infrared (8–14 μm) and lacks cross-band synergy with visible or mid-wave infrared (3–5 μm) spectra. Future directions include: Developing lightweight Transformer-based decision networks integrated with multimodal sensors for real-time environmental perception and millisecond-level dynamic control via reinforcement learning; Combining thermochromic materials for visible-infrared dual-band perturbations to construct multispectral cooperative attack systems; Leveraging physics-informed GANs to simulate extreme thermal noise scenarios and meta-learning frameworks to improve patch adaptability.

Adv-Camou provides a groundbreaking solution for infrared adversarial camouflage, with its dynamic response and cross-scene robustness significantly enhancing battlefield survivability. Through deeper integration of algorithms, materials, and hardware, future advancements could overcome real-time and multi-target adaptability challenges, paving the way for new research pathways in infrared detection countermeasures.

Photographs and biographies of other authors are not available at the time of publication.